\documentstyle[amsfonts,aps]{revtex}

\begin{document}
\title{Density of a gas of spin polarized fermions in a magnetic field}
\author{S. Foulon, F. Brosens, J.\ T. Devreese\thanks{%
Also at Universiteit Antwerpen (RUCA), and at Technische Universiteit
Eindhoven, NL 5600 MB Eindhoven, The Netherlands.}}
\address{Departement Natuurkunde, Universiteit Antwerpen (UIA), Universiteitsplein 1,%
\\
B-2610 Antwerpen}
\author{L. F. Lemmens}
\address{Departement Natuurkunde, Universiteit Antwerpen (RUCA),\\
Groenenborgerlaan171, \\
B-2020 Antwerpen}
\date{To appear in Phys. Rev. E on December 1, 2000}
\maketitle

\begin{abstract}
For a fermion gas with equally spaced energy levels that is subjected to a
magnetic field, the particle density is calculated. The derivation is based
on the path integral approach for identical particles, in combination with
the inversion techniques for the generating function of the static response
functions. Explicit results are presented for the ground state density as a
function of the magnetic field with a number of particles ranging from 1 to
45.
\end{abstract}


\section{Introduction}

The explosive growth of mesoscopic physics in particular enabled to obtain a
tunable number of electrons confined in semiconductor quantum dots \cite
{Kouwenhoven}. Experiments on vertical quantum dots indicated that the
confinement potential of single quantum dots is well described by a
parabolic potential \cite{Tarucha1}. Although the electronic eigenstates and
eigenvalues are essentially given by the Fock-Darwin states \cite
{Fock-Darwin}, detailed experimental data \cite{Ashoori1,Ashoori2,Tarucha2}
reveal the importance of correlation effects on the ground state properties
of the electrons in quantum dots. In order to take these correlation effects
into account, various approximate theoretical methods \cite
{Koskinen,Muller,Lee} have been used. E.g., the eigenstates and eigenvalues
of an harmonic interaction model including the effects of a magnetic field
have been studied with operator techniques \cite{Johnson}, and the harmonic
interaction model has been used to explain specific features in the addition
spectrum\ of a quantum dot in a magnetic field \cite{Angelucci}.

The harmonic interaction model is one of the rare examples for which the
thermodynamical properties are exactly soluble, including the boson or
fermion statistics, in the presence of a magnetic field \cite{Foulon1}. It
can also function as a trial model for the variational treatment of systems
with more realistic interactions with the aid of the Jensen-Feynman
inequality \cite{Feynman2}. An example of this approach can be found in Ref. 
\cite{Tempere}, where the Jensen-Feynman variational approach is used to
describe the Bose-Einstein condensation\ in a gas of $^{87}$Rb and in a gas
of $^{7}$Li atoms. The spin statistics of the harmonic interaction model 
\cite{Spinorgas} can be treated within the same many-body path integral
formalism.

In the present paper, we study the density of harmonically interacting
electrons in a parabolic quantum dot in a magnetic field, taking into
account the electron correlation effects analytically. This analysis is a
natural extension of our previous investigation of the thermodynamical
properties of a confined system of spin-polarized fermions in the presence
of a magnetic field \cite{Foulon1}, using a method that combines the
path-integral formalism \cite{Feynman1}, the method of symmetrized density
matrices \cite{Feynman2}, and inversion techniques for generating functions 
\cite{PRE bosonen,PRE fermionen,LBDSSC99}. Instead of using the stochastic
approach \cite{Simon} with the It\^{o} condition on the magnetic field \cite
{Shulman} for calculating the path integral for $N$ identical interacting
oscillators in a magnetic field, we relied on a detailed investigation of
the classical equations of motion. The quantum mechanical corrections to
their classical action are exactly taken into account.

The model system of $N$ harmonically interacting oscillators in a magnetic
field is described by the Lagrangian (units with $\hbar $ and $m$ equal to
unity are used throughout this paper) 
\begin{equation}
L=\frac{1}{2}\sum_{j=1}^{N}\left( {\bf \dot{r}}_{j}^{2}-2\omega _{c}x_{j}%
\dot{y}_{j}\right) -V,
\end{equation}
where $\omega _{c}$ is the cyclotron frequency, and where the potential
energy $V$ results from a harmonic confinement potential and a harmonic
two-body interaction 
\begin{equation}
V=\frac{\Omega ^{2}}{2}\sum_{j=1}^{N}{\bf r}_{j}^{2}\pm \frac{\omega ^{2}}{4}%
\sum_{j,l=1}^{N}\left( {\bf r}_{j}-{\bf r}_{l}\right) ^{2}.
\end{equation}
The two-body potential might be either attractive or repulsive depending on
the plus sign or the minus sign in the two-body interaction.

As a first step in the calculation of the path integral, we studied the case
of {\em distinguishable} particles \cite{Foulon1}. Because the magnetic
field only affects the equations of motion in the $xy$ plane perpendicular
to the magnetic field, the Lagrangian naturally decouples into two
contributions $L=L_{xy}+L_{z},$ where $L_{xy}$ contains the magnetic field.
This allows to calculate the propagator in the $xy$-plane independently from
the propagator in the $z$-direction. Through a transformation to the center
of mass reference frame, one obtains a set of $N$ three-dimensional
oscillators in a magnetic field. The center of mass is described by an
oscillator with frequency $\Omega .$ The remaining $N-1$ oscillators,
associated with the internal degrees of freedom, have a frequency 
\begin{equation}
w=\sqrt{\Omega ^{2}\pm N\omega ^{2}}.
\end{equation}
Note that the case of a repulsive interaction imposes a stability constraint
on the confinement potential: $\Omega $ has to be large enough to keep the
repelling particles together. The magnetic field does not affect these
frequencies in the $z$ direction. But in the $xy$ plane the equations of
motion reveal two renormalized frequencies due to the magnetic field 
\begin{equation}
s=\sqrt{w^{2}+\omega _{L}^{2}}\text{ \quad and \quad }s_{c.m.}=\sqrt{\Omega
^{2}+\omega _{L}^{2}},
\end{equation}
where $\omega _{L}=\omega _{c}/2$ denotes the Larmor frequency. The index $%
c.m.$ refers to the renormalized frequency in the center of mass coordinates.

Once the propagator for distinguishable particles is known, it is projected
on the antisymmetric representation to obtain the {\em fermion} propagator 
\cite{Foulon1}.

This paper is organized as follows. In Sec. II the one-particle correlation
function is calculated for identical harmonically interacting oscillators in
a homogeneous magnetic field. In Sec. III special attention is paid to the
fermion ground state density in the $xy$- and $xz$-plane. In section IV some
concluding remarks are given.

\section{Static response properties of the model system}

In the path-integral approach to quantum mechanics the expectation value of
a an expression $A\left( {\bf \bar{r}}\right) $ is given by 
\begin{equation}
\left\langle A\left( {\bf \bar{r}},\beta \right) \right\rangle _{I}=\frac{%
\int d{\bf r}_{1}\cdots \int d{\bf r}_{N}K_{I}\left( \left. {\bf \bar{r}}%
,\beta \right| {\bf \bar{r}},0\right) A\left( {\bf \bar{r}}\right) }{\int d%
{\bf r}_{1}\cdots \int d{\bf r}_{N}K_{I}\left( \left. {\bf \bar{r},}\beta
\right| {\bf \bar{r}},0\right) },
\end{equation}
where ${\bf \bar{r}}$ is the $3N$-dimensional vector containing the
coordinates ${\bf r}_{1},\ldots ,{\bf r}_{N}$ of all the $N$ particles, and $%
K_{I}\left( \left. {\bf \bar{r},}\beta \right| {\bf \bar{r}}^{\prime
},0\right) $ denotes the propagator in the Euclidean time $\beta =1/\left(
k_{B}T\right) ,$ with $k_{B}$ denoting the Boltzmann constant and $T$ the
temperature. The subscript $I$ emphasizes that identical particles (fermions
or bosons) are considered. For the probability density and its Fourier
transform this gives 
\begin{equation}
n\left( {\bf r}\right) =\frac{1}{N}\left\langle \sum_{l=1}^{N}\delta \left( 
{\bf r-r}_{l}\right) \right\rangle _{I}=\int \frac{d{\bf q}}{\left( 2\pi
\right) ^{3}}n_{{\bf q}}e^{-i{\bf q\cdot r}}\longleftrightarrow n_{{\bf q}}=%
\frac{1}{N}\sum_{l=1}^{N}\left\langle e^{i{\bf q\cdot r}_{l}}\right\rangle
_{I}.  \label{density and its fouriertransform}
\end{equation}
Substituting the expression for the propagator $K_{I}\left( \left. {\bf \bar{%
r},}\beta \right| {\bf \bar{r}},0\right) $, one obtains the following
integral for the Fourier transform $n_{{\bf q}}$ 
\begin{eqnarray}
n_{{\bf q}} &=&\frac{1}{NZ_{I}\left( \beta ,N\right) }\int \int \frac{d{\bf R%
}d{\bf k}}{\left( 2\pi \right) ^{3}}e^{i{\bf k\cdot R}}\frac{K_{\Omega
}\left( \left. \sqrt{N}Z,\beta \right| \sqrt{N}Z,0\right) K_{\omega
_{L},s_{c.m.}}\left( \left. \sqrt{N}X,\sqrt{N}Y,\beta \right| \sqrt{N}X,%
\sqrt{N}Y,0\right) }{K_{w}\left( \left. \sqrt{N}Z,\beta \right| \sqrt{N}%
Z,0\right) K_{\omega _{L},s}\left( \left. \sqrt{N}X,\sqrt{N}Y,\beta \right| 
\sqrt{N}X,\sqrt{N}Y,0\right) }  \nonumber \\
&&\times \int d{\bf \bar{r}}e^{-i{\bf \bar{k}\cdot \bar{r}}}\sum_{l}e^{i{\bf %
q\cdot r}_{l}}\frac{1}{N!}\sum_{p}\xi ^{p}\prod_{j=1}^{N}K_{w}\left( \left.
\left( Pz\right) _{j},\beta \right| z_{j},0\right) K_{\omega _{L},s}\left(
\left. \left( Px\right) _{j},\left( Py\right) _{j},\beta \right|
x_{j},y_{j},0\right) ,
\end{eqnarray}
where $Z_{I}\left( \beta ,N\right) =\int d{\bf \bar{r}}K_{I}\left( {\bf \bar{%
r}},\beta |{\bf \bar{r}},0\right) $ is the partition function of $N$
identical particles, $P$ denotes a permutation of the particle indices, and $%
\xi =+1$ for bosons and $\xi =-1$ for fermions. The propagators $K_{\Omega
}, $ $K_{w},$ $K_{\omega _{L},s_{c.m.}}$ and $K_{\omega _{L},s}$ are known
in closed form, and explicitly calculated in Ref. \cite{Foulon1}. In order
to obtain tractable expressions for $n_{{\bf q}}$, the summation over all
possible permutations will be rewritten as a sum over all possible cycles.

\subsection{One particle expectation values}

For the one point correlation function a factor $e^{i{\bf q\cdot r}_{l}}$
has to be taken into account in each permutation when applying the cyclic
decomposition of the permutations. Indicating the number of cycles of length 
$\ell $ by $M_{\ell }$, the cyclic decomposition for $n_{{\bf q}}$ becomes 
\begin{eqnarray}
n_{{\bf q}} &=&\frac{1}{NZ_{I}\left( \beta ,N\right) }\int \int \frac{d{\bf R%
}d{\bf k}}{\left( 2\pi \right) ^{3}}e^{i{\bf k\cdot R}}\frac{K_{\Omega
}\left( \left. \sqrt{N}Z,\beta \right| \sqrt{N}Z,0\right) K_{\omega
_{L},s_{c.m.}}\left( \left. \sqrt{N}X,\sqrt{N}Y,\beta \right| \sqrt{N}X,%
\sqrt{N}Y,0\right) }{K_{w}\left( \left. \sqrt{N}Z,\beta \right| \sqrt{N}%
Z,0\right) K_{\omega _{L},s}\left( \left. \sqrt{N}X,\sqrt{N}Y,\beta \right| 
\sqrt{N}X,\sqrt{N}Y,0\right) }  \nonumber \\
&&\times \sum_{M_{1}\ldots M_{N}}\sum_{\ell }\ell M_{\ell }{\cal K}_{\ell
}\left( {\bf k,q}\right) \frac{\xi ^{\left( \ell -1\right) M_{\ell }}}{%
M_{\ell }!\ell ^{M_{\ell }}}\left( {\cal K}_{\ell }\left( {\bf k}\right)
\right) ^{M_{\ell }-1}\prod_{\ell ^{\prime }\neq \ell }\frac{\xi ^{\left(
\ell ^{\prime }-1\right) M_{\ell ^{\prime }}}}{M_{\ell ^{\prime }}!\left(
\ell ^{\prime }\right) ^{M_{\ell ^{\prime }}}}\left( {\cal K}_{\ell ^{\prime
}}\left( {\bf k}\right) \right) ^{M_{\ell ^{\prime }}},
\end{eqnarray}
with 
\begin{equation}
{\cal K}_{\ell }\left( {\bf k,q}\right) =\int d{\bf r}_{\ell +1}\ldots \int d%
{\bf r}_{1}\delta \left( {\bf r}_{\ell +1}-{\bf r}_{1}\right) e^{i{\bf %
q\cdot r}_{1}}\prod_{j=1}^{\ell }K_{\omega _{L,}s}\left( \left.
x_{j+1},y_{j+1},\beta \right| x_{j},y_{j},0\right) K_{w}\left( \left.
z_{j+1},\beta \right| z_{j},0\right) e^{-i{\bf k\cdot r}_{j}/N},
\label{party uno}
\end{equation}
and ${\cal K}_{\ell }\left( {\bf k,q}=0\right) ={\cal K}_{\ell }\left( {\bf k%
}\right) $ is the same function as Eq. (17) of Ref. \cite{Foulon1} in the
calculation of the partition function $Z_{I}\left( \beta ,N\right) $. We
point out that the positive integers $\ell $ and $M_{\ell }$ (with $1\leq
\ell \leq N$) have to satisfy the constraint $\sum_{\ell }\ell M_{\ell }=N$.
Taking into account the semigroup property of the propagators $K_{\omega
_{L},s}\left( x_{j+1},y_{j+1},\beta \mid x_{j},y_{j},0\right) $ and $%
K_{w}\left( \left. z_{j+1},\beta \right| z_{j},0\right) $, one recognizes in 
${\cal K}_{\ell }\left( {\bf k,q}\right) $ the partition function (over a
time interval $\ell \beta $) of a driven harmonic oscillator in a magnetic
field 
\begin{equation}
{\cal K}_{\ell }\left( {\bf k,q}\right) =\int \int \int K_{\omega
_{L,}s}\left( \left. x,y,\ell \beta \right| x,y,0\right) K_{w}\left( \left.
z,\ell \beta \right| z,0\right) e^{-\int_{0}^{\ell \beta }{\bf f}_{{\bf q}%
}\left( \tau \right) {\bf r}\left( \tau \right) d\tau }dxdydz,
\end{equation}
with the driving force 
\begin{equation}
{\bf f}_{{\bf q}}\left( \tau \right) =i\frac{{\bf k}}{N}\sum_{j=0}^{\ell
-1}\delta \left( \tau -j\beta \right) -i{\bf q}\delta \left( \tau \right) .
\label{drivingforce density}
\end{equation}
This partition function is known in closed form and given by 
\begin{equation}
{\cal K}_{\ell }\left( {\bf k,q}\right) =\frac{1}{2\left( \cosh \ell \beta
s-\cosh \ell \beta \omega _{L}\right) }\exp \left( \frac{^{\phi
_{q_{x},q_{y}}}}{4s\left( \cosh \ell \beta s-\cosh \ell \beta \omega
_{L}\right) }\right) \left( \frac{1}{2\sinh \frac{\ell \beta w}{2}}e^{\phi
_{q_{z}}}\right) ,
\end{equation}
with 
\begin{eqnarray}
\phi _{q_{x},q_{y}} &=&\int_{0}^{\ell \beta }\int_{0}^{\ell \beta }\left(
f_{q_{x}}\left( \tau \right) f_{q_{x}}\left( \sigma \right) +f_{q_{y}}\left(
\tau \right) f_{q_{y}}\left( \sigma \right) \right) \left( 
\begin{array}{c}
\cosh \omega _{L}\left( \tau -\sigma \right) \sinh s\left( \ell \beta
-\left| \tau -\sigma \right| \right) \\ 
+\cosh \omega _{L}\left( \ell \beta -\left| \tau -\sigma \right| \right)
\sinh s\left| \tau -\sigma \right|
\end{array}
\right) d\sigma d\tau  \nonumber \\
&&+i\int_{0}^{\ell \beta }\int_{0}^{\ell \beta }\left( f_{q_{x}}\left( \tau
\right) f_{q_{y}}\left( \sigma \right) -f_{q_{x}}\left( \sigma \right)
f_{q_{y}}\left( \tau \right) \right) \left( 
\begin{array}{c}
\sinh \omega _{L}\left( \tau -\sigma \right) \sinh s\left( \ell \beta
-\left| \sigma -\tau \right| \right) \\ 
-\sinh \omega _{L}\left( \ell \beta -\left| \sigma -\tau \right| \right)
\sinh s\left( \tau -\sigma \right)
\end{array}
\right) d\sigma d\tau , \\
\phi _{q_{z}} &=&\frac{1}{2}\int_{0}^{\ell \beta }\int_{0}^{\ell \beta }%
\frac{f_{q_{z}}\left( \tau \right) f_{q_{z}}\left( \sigma \right) }{2w}\frac{%
\cosh \left( \left( \frac{\ell \beta }{2}-\left| \tau -\sigma \right|
w\right) \right) }{\sinh \frac{\ell \beta w}{2}}d\sigma d\tau .
\end{eqnarray}
Substituting the force ${\bf f}_{{\bf q}}\left( \tau \right) $ from (\ref
{drivingforce density}) into the above expressions for $\phi _{q_{x},q_{y}}$
and $\phi _{q_{z}}$ yields 
\begin{eqnarray}
{\cal K}_{\ell }\left( {\bf k,q}\right) &=&{\cal K}_{\ell }\left(
k_{x},k_{y}\right) \exp \left( \frac{\left( k_{x}q_{x}+k_{y}q_{y}\right)
\sinh \beta s}{2Ns\left( \cosh \beta s-\cosh \beta \omega _{L}\right) }-%
\frac{\left( q_{x}^{2}+q_{y}^{2}\right) \sinh \ell \beta s}{4s\left( \cosh
\ell \beta s-\cosh \ell \beta \omega _{L}\right) }\right)  \nonumber \\
&&\times {\cal K}_{\ell }\left( k_{z}\right) \exp \left( \frac{k_{z}q_{z}}{%
2Nw}\coth \frac{1}{2}\beta w-\frac{q_{z}^{2}}{4w}\coth \frac{1}{2}\ell \beta
w\right) ,
\end{eqnarray}
with 
\begin{eqnarray*}
{\cal K}_{\ell }\left( k_{x},k_{y}\right) &=&\frac{1}{2\left( \cosh \ell
\beta s-\cosh \ell \beta \omega _{L}\right) }\exp \left( -\frac{\ell \left(
k_{x}^{2}+k_{y}^{2}\right) }{4N^{2}s}\frac{\sinh \beta s}{\cosh \beta
s-\cosh \beta \omega _{L}}\right) , \\
{\cal K}_{\ell }\left( k_{z}\right) &=&\frac{1}{2\left( \cosh \ell \beta
w-1\right) }\exp \left( -\frac{\ell k_{z}^{2}}{4N^{2}w}\coth \frac{\beta w}{2%
}\right) .
\end{eqnarray*}
The remaining integrations over ${\bf k}$ and ${\bf R}$ in $n_{{\bf q}}$ are
Gaussian and easy to perform, eventually leading to 
\begin{equation}
n_{{\bf q}}=\exp \left( 
\begin{array}{c}
-\frac{\left( q_{x}^{2}+q_{y}^{2}\right) }{4N}\left( \frac{\sinh \beta
s_{c.m.}}{s_{c.m.}\left( \cosh \beta s_{c.m.}-\cosh \beta \omega _{L}\right) 
}-\frac{\sinh \beta s}{s\left( \cosh \beta s-\cosh \beta \omega _{L}\right) }%
\right) \\ 
-\frac{q_{z}^{2}}{4N}\left( \frac{\coth \frac{\beta \Omega }{2}}{\Omega }-%
\frac{\coth \frac{\beta w}{2}}{w}\right)
\end{array}
\right) \tilde{n}_{{\bf q}},
\end{equation}
with 
\begin{eqnarray}
\tilde{n}_{{\bf q}} &=&\frac{1}{N{\Bbb Z}_{I}\left( \beta ,N\right) }%
\sum_{M_{1}\ldots M_{N}}\left[ \sum_{\ell }\ell M_{\ell }\exp \left( -\frac{%
\left( q_{x}^{2}+q_{y}^{2}\right) \sinh \ell \beta s}{4s\left( \cosh \ell
\beta s-\cosh \ell \beta \omega _{L}\right) }-\frac{q_{z}^{2}}{4w}\coth 
\frac{1}{2}\ell \beta w\right) \right]  \nonumber \\
&&\times \prod_{\ell }\frac{\xi ^{\left( \ell -1\right) M_{\ell }}}{M_{\ell
}!\ell ^{M_{\ell }}}\left( \frac{1}{4\left( \cosh \ell \beta s-\cosh \ell
\beta \omega _{L}\right) \sinh \frac{1}{2}\ell \beta w}\right) ^{M_{\ell }}.
\end{eqnarray}
The exponential factor in $n_{{\bf q}}$ accounts for the center of mass
contribution, and it becomes unity for non-interacting particles ($w=\Omega $%
). The factor $\tilde{n}_{{\bf q}}$ is the expectation value of $\sum_{l}e^{i%
{\bf q\cdot r}_{l}}$ in the subspace of the relative coordinate system with
its corresponding partition function ${\Bbb Z}_{I}\left( \beta ,N\right) $.

We now introduce the generating function ${\cal G}_{1}\left( \beta ,u,{\bf q}%
\right) =\sum_{N=0}^{\infty }\left( N{\Bbb Z}_{I}\left( \beta ,N\right) 
\tilde{n}_{{\bf q}}\right) u^{N}$ for the Fourier transform of the density,
as was done before \cite{PRE bosonen,PRE fermionen} in the absence of a
magnetic field, 
\begin{eqnarray}
{\cal G}_{1}\left( \beta ,u,{\bf q}\right) &=&\sum_{N=0}^{\infty
}\sum_{M_{1}\ldots M_{N}}\left[ \sum_{\ell }\ell M_{\ell }\exp \left( -\frac{%
\left( q_{x}^{2}+q_{y}^{2}\right) \sinh \ell \beta s}{8s\sinh \frac{\ell
\beta }{2}\left( s+\omega _{L}\right) \sinh \frac{\ell \beta }{2}\left(
s-\omega _{L}\right) }-\frac{q_{z}^{2}}{4w}\coth \frac{1}{2}\ell \beta
w\right) \right]  \nonumber \\
&&\times \prod_{\ell }\frac{1}{M_{\ell }!}\left( \frac{\xi ^{\left( \ell
-1\right) }u^{\ell }}{\ell \left( 8\sinh \frac{\ell \beta }{2}\left(
s+\omega _{L}\right) \sinh \frac{\ell \beta }{2}\left( s-\omega _{L}\right)
\sinh \frac{1}{2}\ell \beta w\right) }\right) ^{M_{\ell }}, \\
&=&\Xi _{I}\left( \beta ,u\right) \sum_{\ell =1}^{\infty }\frac{\xi ^{\left(
\ell -1\right) }\exp \left( -\frac{\left( q_{x}^{2}+q_{y}^{2}\right) \sinh
\ell \beta s}{8s\sinh \frac{\ell \beta }{2}\left( s+\omega _{L}\right) \sinh 
\frac{\ell \beta }{2}\left( s-\omega _{L}\right) }-\frac{q_{z}^{2}}{4w}\coth 
\frac{1}{2}\ell \beta w\right) }{8\sinh \frac{\ell \beta }{2}\left( s+\omega
_{L}\right) \sinh \frac{\ell \beta }{2}\left( s-\omega _{L}\right) \sinh 
\frac{1}{2}\ell \beta w}u^{\ell },  \label{generating function}
\end{eqnarray}
where $\Xi _{I}\left( \beta ,u\right) =\sum_{N=0}^{\infty }{\Bbb Z}%
_{I}\left( \beta ,N\right) u^{N}$ is the generating function for the
partition function ${\Bbb Z}_{I}\left( \beta ,N\right) $. After
straightforward algebra one is left with 
\begin{equation}
\tilde{n}_{{\bf q}}=\frac{1}{N}\sum_{\ell =1}^{N}\xi ^{\left( \ell -1\right)
}\frac{{\Bbb Z}_{I}\left( \beta ,N-\ell \right) }{{\Bbb Z}_{I}\left( \beta
,N\right) }\frac{\exp \left( -\frac{\left( q_{x}^{2}+q_{y}^{2}\right) \sinh
\ell \beta s}{8s\sinh \frac{\ell \beta }{2}\left( s+\omega _{L}\right) \sinh 
\frac{\ell \beta }{2}\left( s-\omega _{L}\right) }-\frac{q_{z}^{2}}{4w}\coth 
\frac{1}{2}\ell \beta w\right) }{8\sinh \frac{\ell \beta }{2}\left( s+\omega
_{L}\right) \sinh \frac{\ell \beta }{2}\left( s-\omega _{L}\right) \sinh 
\frac{1}{2}\ell \beta w}.
\end{equation}
It is noted that in the limit ${\bf q}\rightarrow 0$ the sum rule $\tilde{n}%
_{{\bf q}=0}=1$ is indeed satisfied. The density $n\left( {\bf r}\right) $
in real space then becomes 
\begin{eqnarray}
n\left( {\bf r}\right) &=&\int \frac{d{\bf q}}{\left( 2\pi \right) ^{3}}n_{%
{\bf q}}e^{-i{\bf q\cdot r}} \\
&=&\frac{1}{N}\sum_{\ell =1}^{N}\xi ^{\left( \ell -1\right) }\frac{{\Bbb Z}%
_{I}\left( \beta ,N-\ell \right) }{{\Bbb Z}_{I}\left( \beta ,N\right) }\frac{%
s{\cal B}_{\ell }}{\pi }\sqrt{\frac{w{\cal A}_{\ell }}{\pi }}\frac{\exp
\left( -s{\cal B}_{\ell }\left( x^{2}+y^{2}\right) -w{\cal A}_{\ell
}z^{2}\right) }{8\sinh \frac{\ell \beta }{2}\left( s+\omega _{L}\right)
\sinh \frac{\ell \beta }{2}\left( s-\omega _{L}\right) \sinh \frac{1}{2}\ell
\beta w},  \label{density with signproblem}
\end{eqnarray}
with 
\begin{eqnarray}
{\cal A}_{\ell } &=&\left[ \coth \frac{1}{2}\ell \beta w+\frac{1}{N}\left( 
\frac{w}{\Omega }\coth \frac{\beta \Omega }{2}-\coth \frac{\beta w}{2}%
\right) \right] ^{-1}, \\
{\cal B}_{\ell } &=&\left[ \frac{\sinh \ell \beta s}{\cosh \ell \beta
s-\cosh \ell \beta \omega _{L}}+\frac{1}{N}\left( \frac{s}{s_{c.m.}}\frac{%
\sinh \beta s_{c.m.}}{\cosh \beta s_{c.m.}-\cosh \beta \omega _{L}}-\frac{%
\sinh \beta s}{\cosh \beta s-\cosh \beta \omega _{L}}\right) \right] ^{-1}.
\end{eqnarray}
The sum rule $\int d{\bf r}n\left( {\bf r}\right) =1$ for the density is
easily verified. In the next subsection the fermion ground state density
will be examined, by inverting the defining series for the generating
function ${\cal G}_{1}\left( \beta ,u,{\bf q}\right) $. Subsequently results
will be presented for the ground state density in the $xy$-plane and in the $%
xz$-plane.

\section{Ground state density}

Because of alternating signs in the recurrence relations for the partition
functions${\Bbb \ Z}_{I}\left( \beta ,N\right) ,$ the equation (\ref{density
with signproblem}) is not appropriate for numerical purposes, in particular
in the low temperature limit. This sign problem can be circumvented by an
alternative inversion of the generating function ${\cal G}_{1}$ using
contour integration: 
\begin{equation}
\tilde{n}_{{\bf q}}=\frac{1}{N!}\frac{1}{N{\Bbb Z}_{I}\left( \beta ,N\right) 
}\left. \frac{\partial ^{N}{\cal G}_{1}\left( \beta ,u,{\bf q}\right) }{%
\partial u^{N}}\right| _{u=0}=\frac{1}{2\pi {\Bbb Z}_{I}\left( \beta
,N\right) }\frac{1}{N}\int_{0}^{2\pi }\frac{{\cal G}_{1}\left( \beta
,ue^{i\theta },{\bf q}\right) }{u^{N}}e^{-iN\theta }d\theta .
\end{equation}
Substituting (\ref{generating function}) into the above expression yields 
\begin{eqnarray}
\tilde{n}_{{\bf q}} &=&\frac{1}{N}\frac{\Xi _{I}\left( \beta ,u\right) /u^{N}%
}{2\pi {\Bbb Z}_{I}\left( \beta ,N\right) }e^{-\left( \kappa _{x}^{2}+\kappa
_{y}^{2}+\kappa _{z}^{2}\right) }  \nonumber \\
&&\times \sum_{\ell =1}^{\infty }\int_{0}^{2\pi }\frac{\Xi _{I}\left( \beta
,ue^{i\theta }\right) }{\Xi _{I}\left( \beta ,u\right) }\frac{\xi ^{\left(
\ell -1\right) }\left( \sqrt{bb_{1}b_{2}}ue^{i\theta }\right) ^{\ell }\exp
\left( -\left( \kappa _{x}^{2}+\kappa _{y}^{2}\right) \left( \frac{%
b_{1}^{\ell }}{_{1-b_{1}^{\ell }}}+\frac{b_{2}^{\ell }}{_{1-b_{2}^{\ell }}}%
\right) \right) \exp \left( -2\kappa _{z}^{2}\frac{b^{\ell }}{_{1-b^{\ell }}}%
\right) }{\left( 1-b^{\ell }\right) \left( 1-b_{1}^{\ell }\right) \left(
1-b_{2}^{\ell }\right) }e^{-iN\theta }d\theta ,
\end{eqnarray}
with the short hand notations 
\begin{equation}
\kappa _{x}^{2}=q_{x}^{2}/4s,\text{\quad }\kappa _{y}^{2}=q_{y}^{2}/4s,\text{%
\quad }\kappa _{z}^{2}=q_{z}^{2}/4w,\text{\quad }b=e^{-\beta w},\text{\quad }%
b_{1}=e^{-\beta \left( s+\omega _{L}\right) },\text{\quad }b_{2}=e^{-\beta
\left( s-\omega _{L}\right) }.
\end{equation}
By expanding the Fourier transform of the density $\tilde{n}_{{\bf q}}$ in
powers of $b$, $b_{1}$ and $b_{2}$ one arrives at 
\begin{eqnarray}
\tilde{n}_{{\bf q}} &=&\frac{1}{N}\frac{\Xi _{I}\left( \beta ,u\right) }{%
2\pi {\Bbb Z}_{I}\left( \beta ,N\right) u^{N}}e^{-\left( \kappa
_{x}^{2}+\kappa _{y}^{2}+\kappa _{z}^{2}\right) }\sum_{m=0}^{\infty
}\sum_{m_{1}=0}^{\infty }\sum_{n_{1}=0}^{\infty }\sum_{m_{2}=0}^{\infty
}\sum_{n_{2}=0}^{\infty }\frac{\left( -\kappa _{x}^{2}\right)
^{m_{1}+m_{2}}\left( -\kappa _{y}^{2}\right) ^{n_{1}+n_{2}}\left( -2\kappa
_{z}^{2}\right) ^{m}}{\Gamma \left( m_{1}+1\right) \Gamma \left(
m_{2}+1\right) \Gamma \left( n_{1}+1\right) \Gamma \left( n_{2}+1\right)
\Gamma \left( m+1\right) }  \nonumber \\
&&\times \sum_{k=0}^{\infty }\sum_{k_{1}=0}^{\infty }\sum_{k_{2}=0}^{\infty }%
\frac{\left( m_{1}+n_{1}+1\right) _{k_{1}}\left( m_{2}+n_{2}+1\right)
_{k_{2}}\left( m+1\right) _{k}}{\Gamma \left( k_{1}+1\right) \Gamma \left(
k_{2}+1\right) \Gamma \left( k+1\right) }  \nonumber \\
&&\times \int_{0}^{2\pi }\frac{ue^{i\theta }b^{m+k+\frac{1}{2}%
}b_{1}^{k_{1}+m_{1}+n_{1}+\frac{1}{2}}b_{2}^{k_{2}+m_{2}+n_{2}+\frac{1}{2}%
}\Psi _{N-1}\left( \theta \right) -\xi \left( ub^{m+k+\frac{1}{2}%
}b_{1}^{k_{1}+m_{1}+n_{1}+\frac{1}{2}}b_{2}^{k_{2}+m_{2}+n_{2}+\frac{1}{2}%
}\right) ^{2}\Psi _{N}\left( \theta \right) }{1-2\xi ub^{m+k+\frac{1}{2}%
}b_{1}^{k_{1}+m_{1}+n_{1}+\frac{1}{2}}b_{2}^{k_{2}+m_{2}+n_{2}+\frac{1}{2}%
}\cos \theta +\left( ub^{m+k+\frac{1}{2}}b_{1}^{k_{1}+m_{1}+n_{1}+\frac{1}{2}%
}b_{2}^{k_{2}+m_{2}+n_{2}+\frac{1}{2}}\right) ^{2}}d\theta ,
\end{eqnarray}
with $\left( a\right) _{p}=\Gamma \left( a+p\right) /\Gamma \left( a\right) $
the Pochhammer symbol. The function $\Psi _{N}\left( \theta \right)
=e^{-iN\theta }\Xi _{I}\left( \beta ,ue^{i\theta }\right) /\Xi _{I}\left(
\beta ,u\right) $ has previously \cite{Foulon1} been obtained. Using (\ref
{density and its fouriertransform}), the density (still at arbitrary
temperature) then becomes 
\begin{eqnarray}
&&n\left( {\bf r}\right) =  \nonumber \\
&&\frac{1}{N}\frac{1}{\int_{0}^{2\pi }\Psi _{N}\left( \theta \right) d\theta 
}\sqrt{\frac{s^{2}w}{\pi ^{3}A^{2}B}}\exp \left( -\frac{s\left(
x^{2}+y^{2}\right) }{A}-\frac{wz^{2}}{B}\right)  \nonumber \\
&&\times \sum_{m=0}^{\infty }\sum_{m_{1}=0}^{\infty }\sum_{n_{1}=0}^{\infty
}\sum_{m_{2}=0}^{\infty }\sum_{n_{2}=0}^{\infty }\frac{\left( -\frac{1}{4A}%
\right) ^{m_{1}+m_{2}+n_{1}+n_{2}}\left( -\frac{1}{2B}\right) ^{m}}{\Gamma
\left( m_{1}+1\right) \Gamma \left( m_{2}+1\right) \Gamma \left(
n_{1}+1\right) \Gamma \left( n_{2}+1\right) \Gamma \left( m+1\right) } 
\nonumber \\
&&\times \sum_{k=0}^{\infty }\sum_{k_{1}=0}^{\infty }\sum_{k_{2}=0}^{\infty }%
\frac{\left( m_{1}+n_{1}+1\right) _{k_{1}}\left( m_{2}+n_{2}+1\right)
_{k_{2}}\left( m+1\right) _{k}}{\Gamma \left( k_{1}+1\right) \Gamma \left(
k_{2}+1\right) \Gamma \left( k+1\right) }  \nonumber \\
&&\times
\sum_{q=0}^{m}\sum_{q_{1}=0}^{m_{1}+m_{2}}\sum_{q_{2}=0}^{n_{1}+n_{2}}\frac{%
\left( -\frac{4sx^{2}}{A}\right) ^{m_{1}+m_{2}-q_{1}}\left( -\frac{4sy^{2}}{A%
}\right) ^{n_{1}+n_{2}-q_{2}}\left( -\frac{4wz^{2}}{B}\right) ^{m-q}}{\left(
2\left( m_{1}+m_{2}\right) +1\right) _{-2q_{1}}\Gamma \left( q_{1}+1\right)
\left( 2\left( n_{1}+n_{2}\right) +1\right) _{-2q_{2}}\Gamma \left(
q_{2}+1\right) \left( 2m+1\right) _{-2q}\Gamma \left( q+1\right) }  \nonumber
\\
&&\times \int_{0}^{2\pi }\frac{ue^{i\theta }b^{m+k+\frac{1}{2}%
}b_{1}^{k_{1}+m_{1}+n_{1}+\frac{1}{2}}b_{2}^{k_{2}+m_{2}+n_{2}+\frac{1}{2}%
}\Psi _{N-1}\left( \theta \right) -\xi \left( ub^{m+k+\frac{1}{2}%
}b_{1}^{k_{1}+m_{1}+n_{1}+\frac{1}{2}}b_{2}^{k_{2}+m_{2}+n_{2}+\frac{1}{2}%
}\right) ^{2}\Psi _{N}\left( \theta \right) }{1-2\xi ub^{m+k+\frac{1}{2}%
}b_{1}^{k_{1}+m_{1}+n_{1}+\frac{1}{2}}b_{2}^{k_{2}+m_{2}+n_{2}+\frac{1}{2}%
}\cos \theta +\left( ub^{m+k+\frac{1}{2}}b_{1}^{k_{1}+m_{1}+n_{1}+\frac{1}{2}%
}b_{2}^{k_{2}+m_{2}+n_{2}+\frac{1}{2}}\right) ^{2}}d\theta ,
\label{density at arbitrary temperature}
\end{eqnarray}
with 
\begin{eqnarray}
A &=&1+\frac{1}{N}\left( \frac{s}{s_{c.m.}}\frac{\sinh \beta s_{c.m.}}{\cosh
\beta s_{c.m.}-\cosh \beta \omega _{L}}-\frac{\sinh \beta s}{\cosh \beta
s-\cosh \beta \omega _{L}}\right) , \\
B &=&1+\frac{1}{N}\left( \frac{w}{\Omega }\coth \frac{\beta \Omega }{2}%
-\coth \frac{\beta w}{2}\right) .
\end{eqnarray}
The ground state density at $T=0$ is then readily obtained by taking the
limit $\beta \rightarrow \infty $: 
\begin{eqnarray}
&&n_{T=0}\left( {\bf r}\right)  \nonumber \\
&=&\frac{1}{N}\sqrt{\frac{\left( \sigma s\right) ^{2}\vartheta w}{\pi ^{3}}}%
\exp \left( -\sigma s\left( x^{2}+y^{2}\right) -\vartheta wz^{2}\right) \\
&&\times \sum_{m=0}^{\infty }\sum_{m_{1}=0}^{\infty }\sum_{n_{1}=0}^{\infty
}\sum_{m_{2}=0}^{\infty }\sum_{n_{2}=0}^{\infty }\frac{\left( -\frac{\sigma 
}{4}\right) ^{m_{1}+m_{2}+n_{1}+n_{2}}\left( -\frac{\vartheta }{2}\right)
^{m}}{\Gamma \left( m_{1}+1\right) \Gamma \left( m_{2}+1\right) \Gamma
\left( n_{1}+1\right) \Gamma \left( n_{2}+1\right) \Gamma \left( m+1\right) }
\nonumber \\
&&\times \sum_{k=0}^{\infty }\sum_{k_{1}=0}^{\infty }\sum_{k_{2}=0}^{\infty }%
\frac{\left( m_{1}+n_{1}+1\right) _{k_{1}}\left( m_{2}+n_{2}+1\right)
_{k_{2}}\left( m+1\right) _{k}}{\Gamma \left( k_{1}+1\right) \Gamma \left(
k_{2}+1\right) \Gamma \left( k+1\right) }  \nonumber \\
&&\times
\sum_{q=0}^{m}\sum_{q_{1}=0}^{m_{1}+m_{2}}\sum_{q_{2}=0}^{n_{1}+n_{2}}\frac{%
\left( -4\sigma sx^{2}\right) ^{m_{1}+m_{2}-q_{1}}\left( -4\sigma
sy^{2}\right) ^{n_{1}+n_{2}-q_{2}}\left( -4\vartheta wz^{2}\right) ^{m-q}}{%
\left( 2\left( m_{1}+m_{2}\right) +1\right) _{-2q_{1}}\Gamma \left(
q_{1}+1\right) \left( 2\left( n_{1}+n_{2}\right) +1\right) _{-2q_{2}}\Gamma
\left( q_{2}+1\right) \left( 2m+1\right) _{-2q}\Gamma \left( q+1\right) },
\label{density at zero temperature}
\end{eqnarray}
with 
\begin{equation}
\sigma =\frac{N}{N-1+s/s_{c.m.}}\quad \text{and\quad }\vartheta =\frac{N}{%
N-1+w/\Omega }.
\end{equation}
The sum rule $\int d{\bf r}n_{T=0}\left( {\bf r}\right) =1$ can easily be
verified.

\subsection{Density in the $xy$-plane}

The ground state surface density $n(x,y)\equiv n_{T=0}\left( x,y,z\right)
/n_{T=0}(0,0,z)$ in the $xy$ plane is cylindrically symmetric, which means
that $n(x=0,y)$ contains all the information. Figure 1 shows $Nn(x=0,y)/w$
for $1$ up to $45$ electrons for the non-interacting case ($w=\Omega $) at
zero temperature. The factor $1/w$ makes this expression dimensionless (in
the units with $\hbar =m=1$ used throughout this paper). The factor $N$ is
introduced for clarity in the figure, in order to avoid too strongly
overlapping curves. The densities for $N=1$, $3$, $6$, $10$, $15$, $21$, $28$%
, $36$, $45$ electrons, i.e. for closed shell configurations, are emphasized
by dashed lines. Oscillations appear in the density profiles, indicating
concentric orbitals of increased density around the center of the
confinement potential. The oscillations become more pronounced if the number
of particles increases. Previously \cite{Tempere qdot} the effects of an
attractive $(w>\Omega )$ and a repulsive $(w<\Omega )$ two-particle
interaction have been studied in the absence of a magnetic field. It was
seen that a repulsive interaction induced an expansion whereas an attractive
interaction induced a contraction of the gas of interacting fermions. These
effects also appear for non-zero magnetic fields, but for brevity reasons
they are not plotted.

With increasing magnetic field the oscillations in the density gradually
become less pronounced and the confinement is enhanced as $\omega _{L}$
increases. The density profile changes whenever the magnetic susceptibility
(which is proportional to $\partial E_{G}/\partial \omega _{L}$ with $%
E_{G}=\sum_{E<E_{F}}E$ the ground state energy) exhibits a discontinuity as
a function of the magnetic field. For $\omega _{L}\gg w$ the energy spectrum
behaves like a Landau spectrum. This is illustrated for the non-interacting
case ($w=\Omega $) in fig. 2a and fig. 2b for $N=10$ electrons, and in fig.
3a and fig. 3b for $N=28$ electrons. The insets show the magnetic
susceptibility as a function of $\omega _{L}$.

\subsection{Density in the $xz$-plane}

Similarly as for the density in the $xy$-plane, the density profile in the $%
xz$-plane changes whenever there is a jump in the magnetic susceptibility.
Figure 4 shows density contour plots for 4 fermions ($w=\Omega $) for
various magnetic fields. The asymmetry of the $x$- and $z$-direction due to
the magnetic field along the $z$-axis is clearly revealed in the plots. In
fig. 5 the density contours for 10 fermions ($w=\Omega $) in the $xz$-plane
are illustrated.

\section{Conclusion and discussion}

In this paper we have presented analytical results for the density of
spin-polarized harmonically interacting fermion oscillators in a magnetic
field, taking the fermion statistics of the particles into account. The
approach presented here is valid for any number of electrons and for any
temperature. We concentrated on the ground state density for a number of
particles ranging from 1 up to 45, although also higher particle numbers can
be treated. The density, in the $xy$- as well as in the $xz$-plane, shows a
magnetic field dependency that is governed by the discontinuity in the
magnetic susceptibility. Oscillations are present in the density and they
are more pronounced as the number of particles is increased. These
oscillations are smoothed out and finally disappear with increasing Larmor
frequency $\omega _{L}$. Whenever the magnetic field causes a discontinuity
in the magnetic susceptibility, the density profile undergoes a sudden
change, thus providing a means for characterizing the parameters of the
system. To the best of our knowledge, the path integral approach used in
this paper is the only method so far that provides this detailed information
on the density for an interacting fermion system.

\acknowledgements%
%

Part of this work is performed in the framework of the FWO projects No.
G.0287.95, 1.5.545.98, G.0071.98, and WO.073.94N [Wetenschappelijke
Onderzoeksgemeenschap over ``Laag-dimensionele systemen'' (Scientific
Research Community on Low-Dimensional Systems)], the ``Interuniversitaire
Attractiepolen -- Belgische Staat, Diensten van de Eerste Minister --
Wetenschappelijke, Technische en Culturele aangelegenheden''
(Interuniversity Poles of Attraction Programs -- Belgian State, Prime
Minister's Office -- Federal Office for Scientific, Technical and Cultural
Affairs), and in the framework of the GOA BOF UA 2000 projects of the
Universiteit Antwerpen. The authors F. Brosens and S. Foulon acknowledge the
FWO (Fonds voor Wetenschappelijk Onderzoek -- Vlaanderen) for financial
support.

\begin{center}
\newpage

{\bf Figure captions}
\end{center}

\begin{description}
\item  {\bf Fig. 1: }Scaled surface density $Nn\left( x=0,y\right) w$ for $%
\Omega =w$ in the plane perpendicular to the magnetic field, as a function
of the scaled distance $y/y_{0}$ (with $y_{0}=1/\sqrt{w}$, units with $\hbar
=m=1$ are used) from the centre of the parabolic confinement potential for $%
N=1,\ldots ,45$ fermions in the limit $\omega _{L}/w\rightarrow 0$. The
densities corresponding to 1, 3, 6, 10, 15, 21, 28, 36, 45 fermions (i.e.
for closed shell configurations in the absence of a magnetic field) are
indicated by dashed lines and the corresponding particle number is
explicitly indicated.

\item  {\bf Fig. 2: a}) Scaled surface density $Nn\left( x=0,y\right) /w$
for $\Omega =w$ in the plane perpendicular to the magnetic field, as a
function of the scaled distance $y/y_{0}$ (with $y_{0}=1/\sqrt{w})$ from the
centre of the parabolic confinement potential for $N=10$ fermions and for
various Larmor frequencies $\omega _{L}/w=0.2,$ $0.6,$ $1.2,$ $3.0$. The
inset shows the magnetic susceptibility for $N=10$ fermions as a function of
the Larmor frequency; {\bf b}) Same as in Fig. 2a, but for $\omega _{L}/w=0,$
$0.4,$ $0.9$.

\item  {\bf Fig. 3: a}) Scaled surface density $Nn\left( x=0,y\right) /w$
for $\Omega =w$ as a function of the scaled distance $y/y_{0}$ (with $%
y_{0}=1/\sqrt{w})$ from the centre of the parabolic confinement potential
for $N=28$ fermions and for various Larmor frequencies $\omega _{L}/w=0,$ $%
1.0,$ $2.0,$ $3.0$. The inset shows the magnetic susceptibility for $N=28$
fermions as a function of $\omega _{L}$; {\bf b)} Same as in Fig. 3a, but
for $\omega _{L}/w=0.5,$ $1.5,$ $2.5,$ $3.5$.

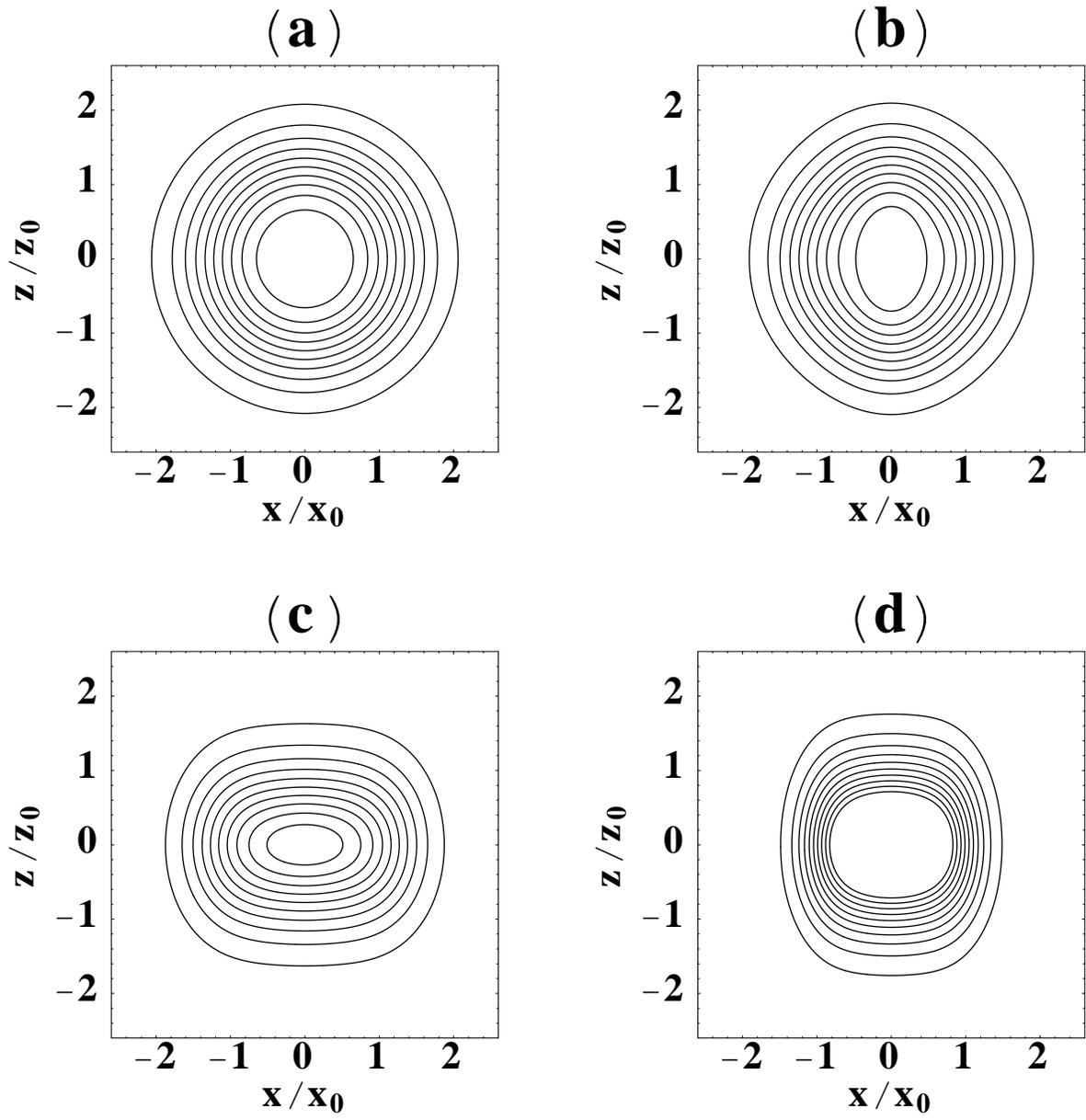
\item  {\bf Fig. 4:} Contourplots of $Nn\left( x,z\right) /w$ for $N=4$
fermions and for $\Omega =w$ as a function of the scaled coordinates $%
x/x_{0} $ and $z/z_{0}$ with $x_{0}=$ $z_{0}=1/\sqrt{w}$, measured from the
center of the parabolic confinement potential, for different values of the
magnetic field: (a) $\omega _{L}/w=0.2,$ (b) $\omega _{L}/w=1.0,$ (c) $%
\omega _{L}/w=1.5,$ and (d) $\omega _{L}/w=3.0$.

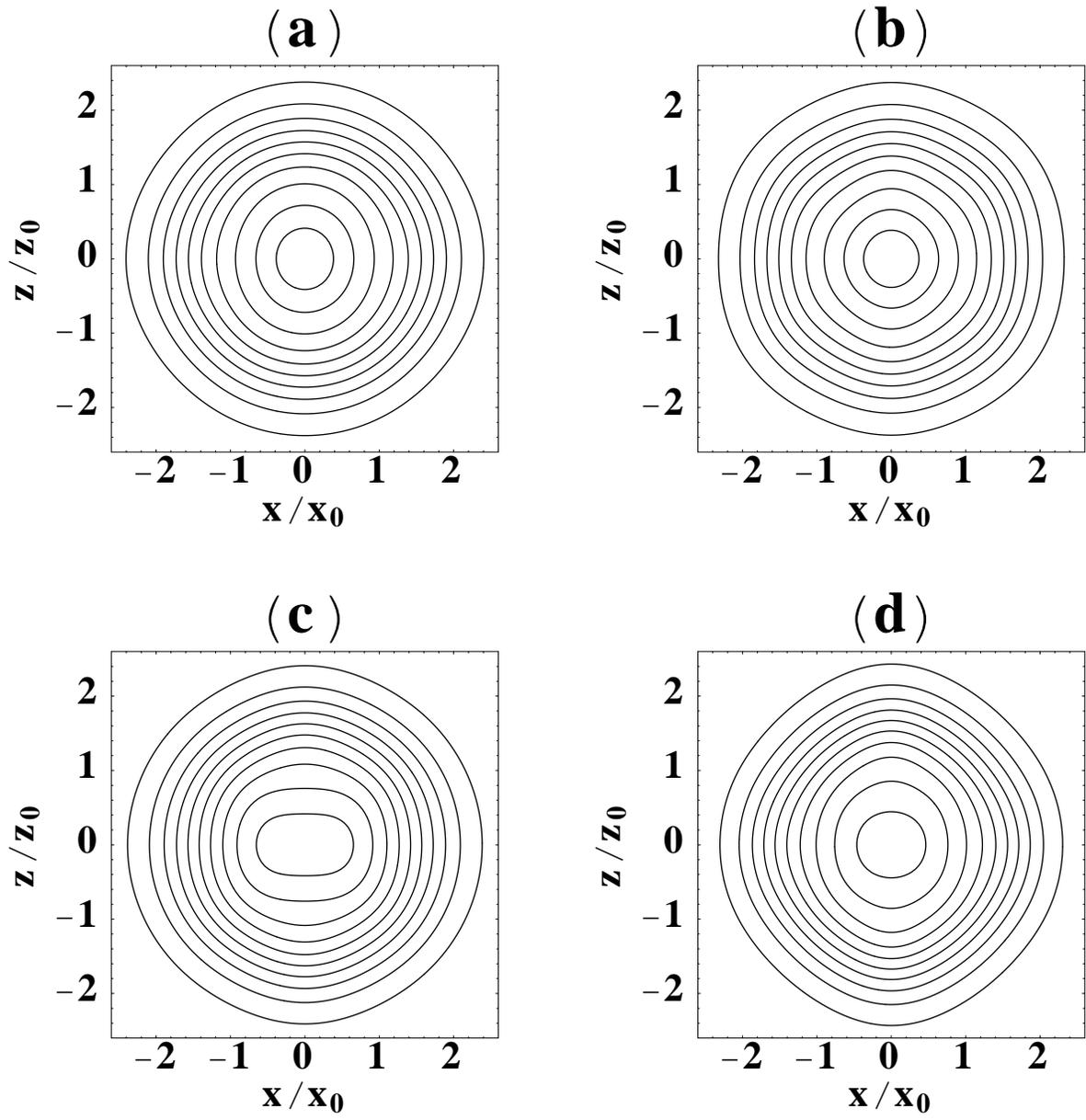
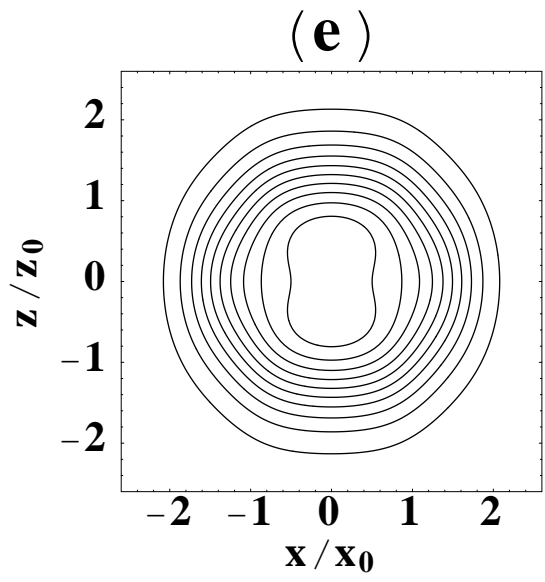
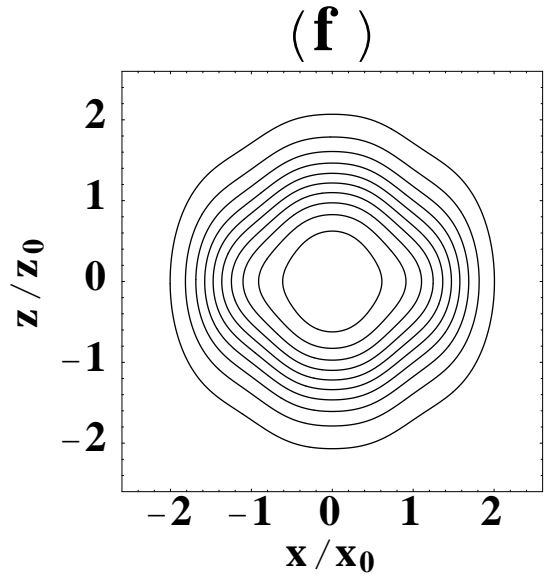
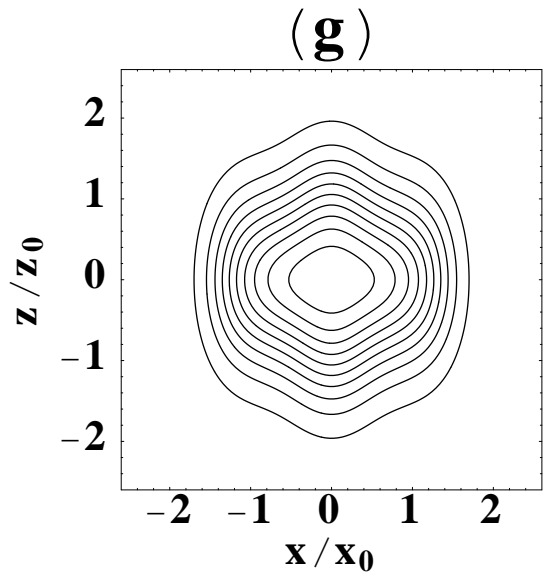
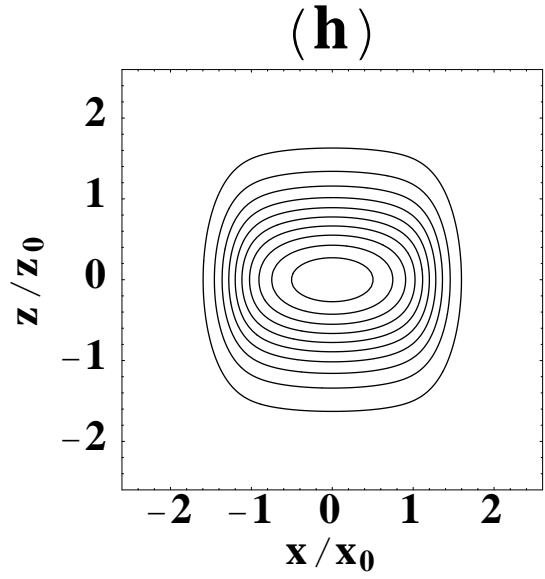
\item  {\bf Fig. 5:} Contourplots of $Nn\left( x,z\right) /w$ for $N=10$
fermions and for $\Omega =w$ as a function of the scaled coordinates $%
x/x_{0} $ and $z/z_{0}$ with $x_{0}=$ $z_{0}=1/\sqrt{w}$, measured from the
center of the parabolic confinement potential for different values of the
magnetic field: (a) $\omega _{L}/w=0.3,$ (b) $\omega _{L}/w=0.5,$ (c) $%
\omega _{L}/w=0.7,$ (d) $\omega _{L}/w=1.2$; (e) $\omega _{L}/w=2.0,$ (f) $%
\omega _{L}/w=2.5,$ (g) $\omega _{L}/w=4.0,$ (h) $\omega _{L}/w=5.0$.
\end{description}

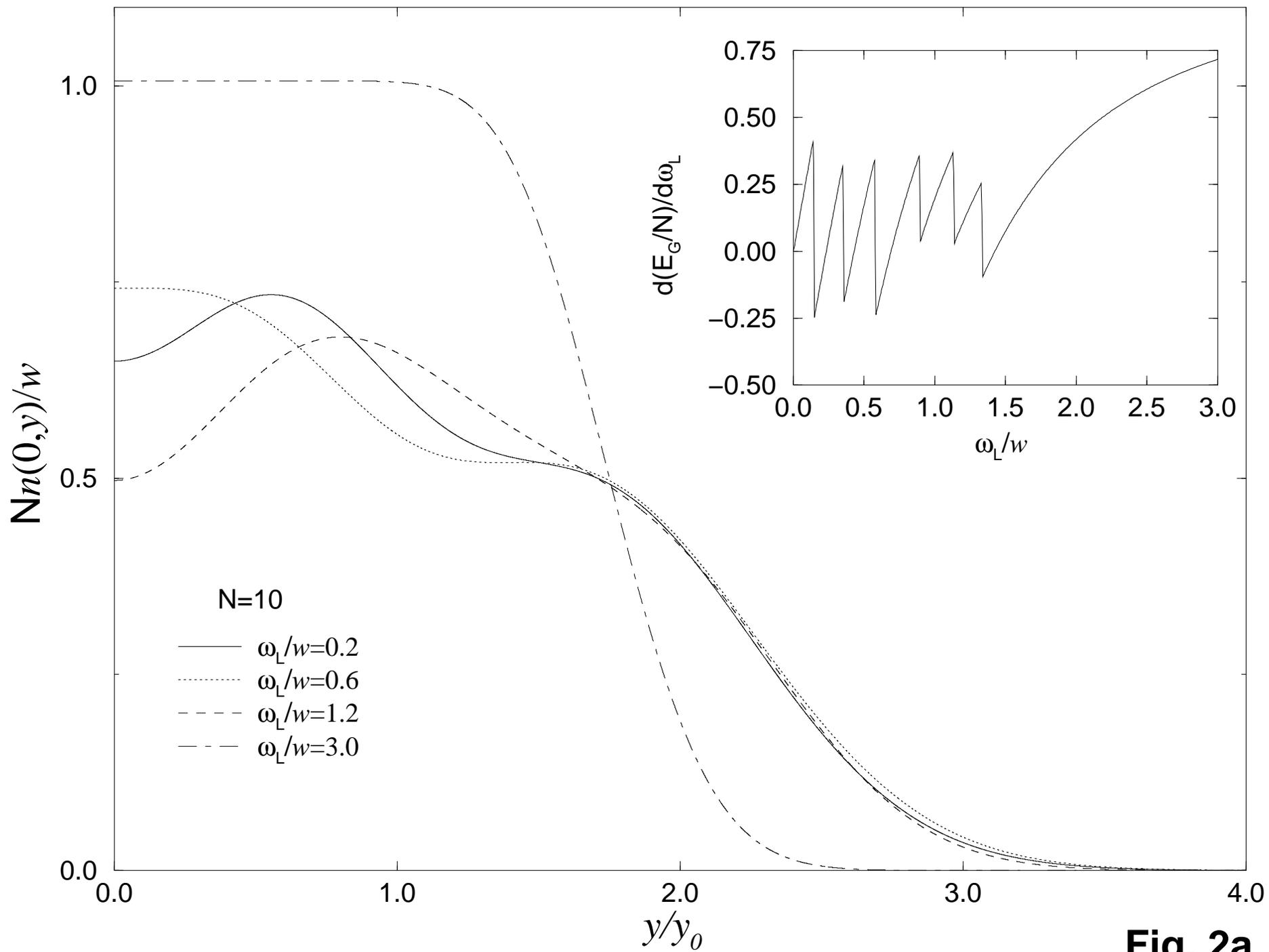
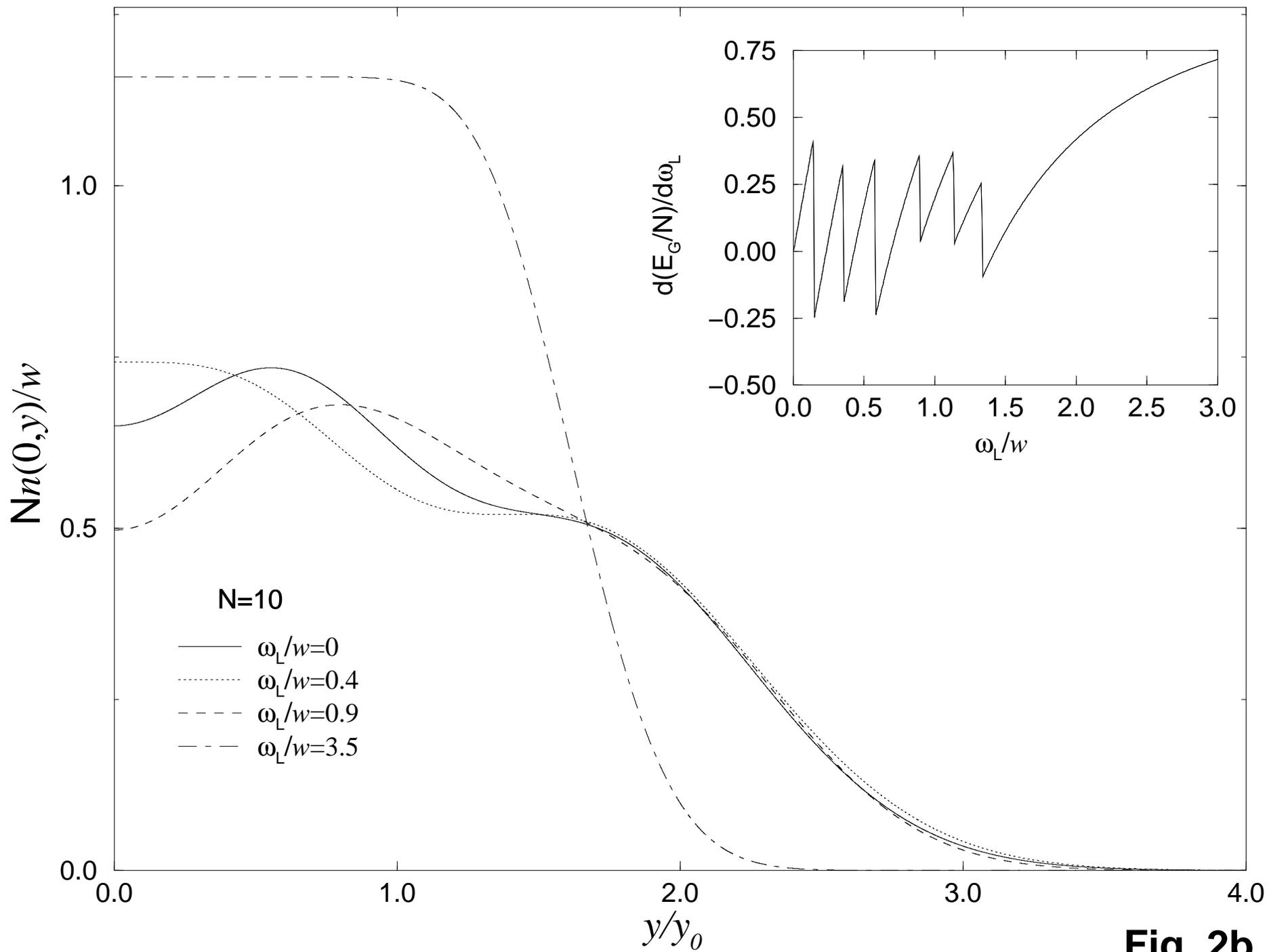
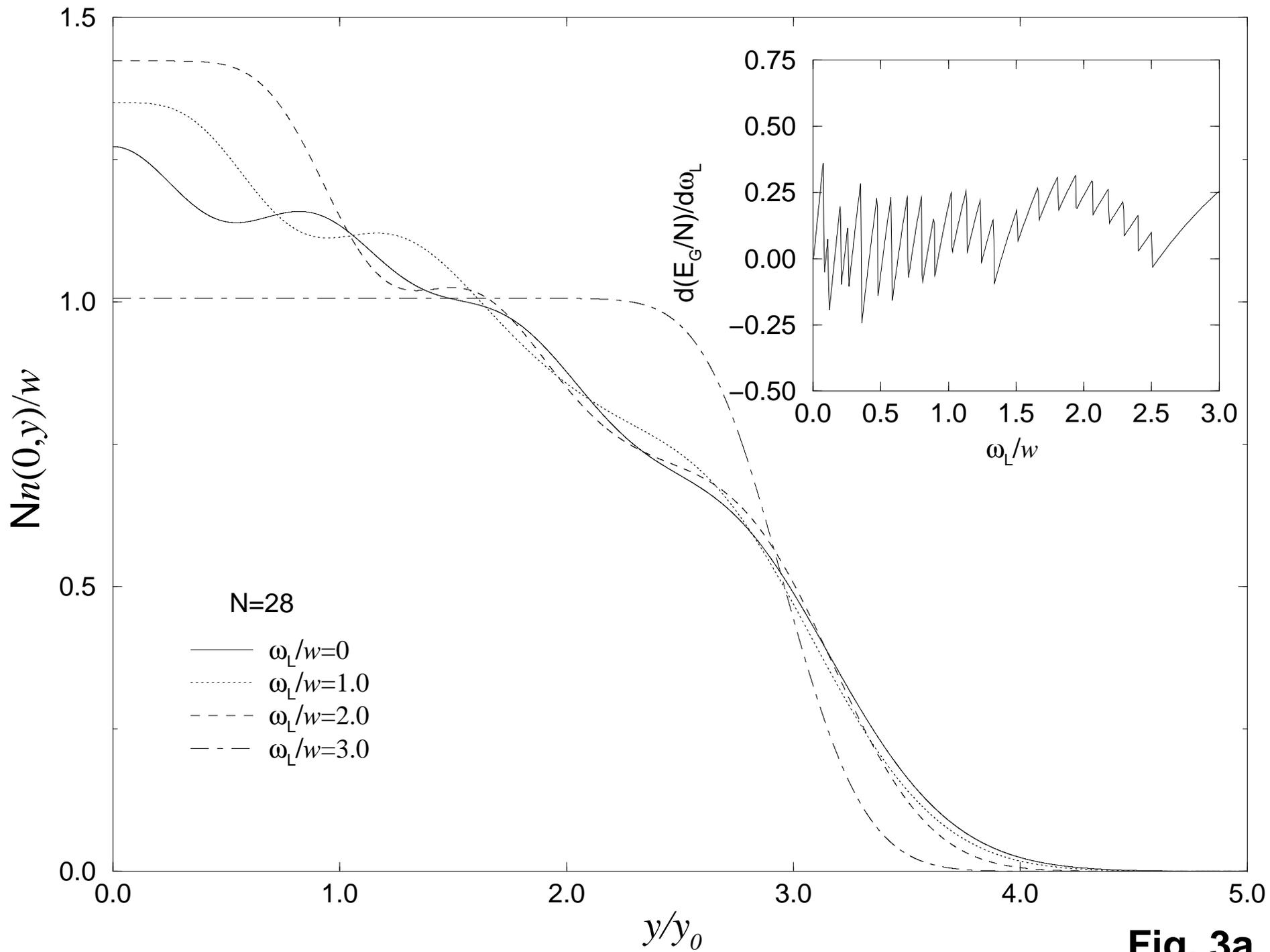
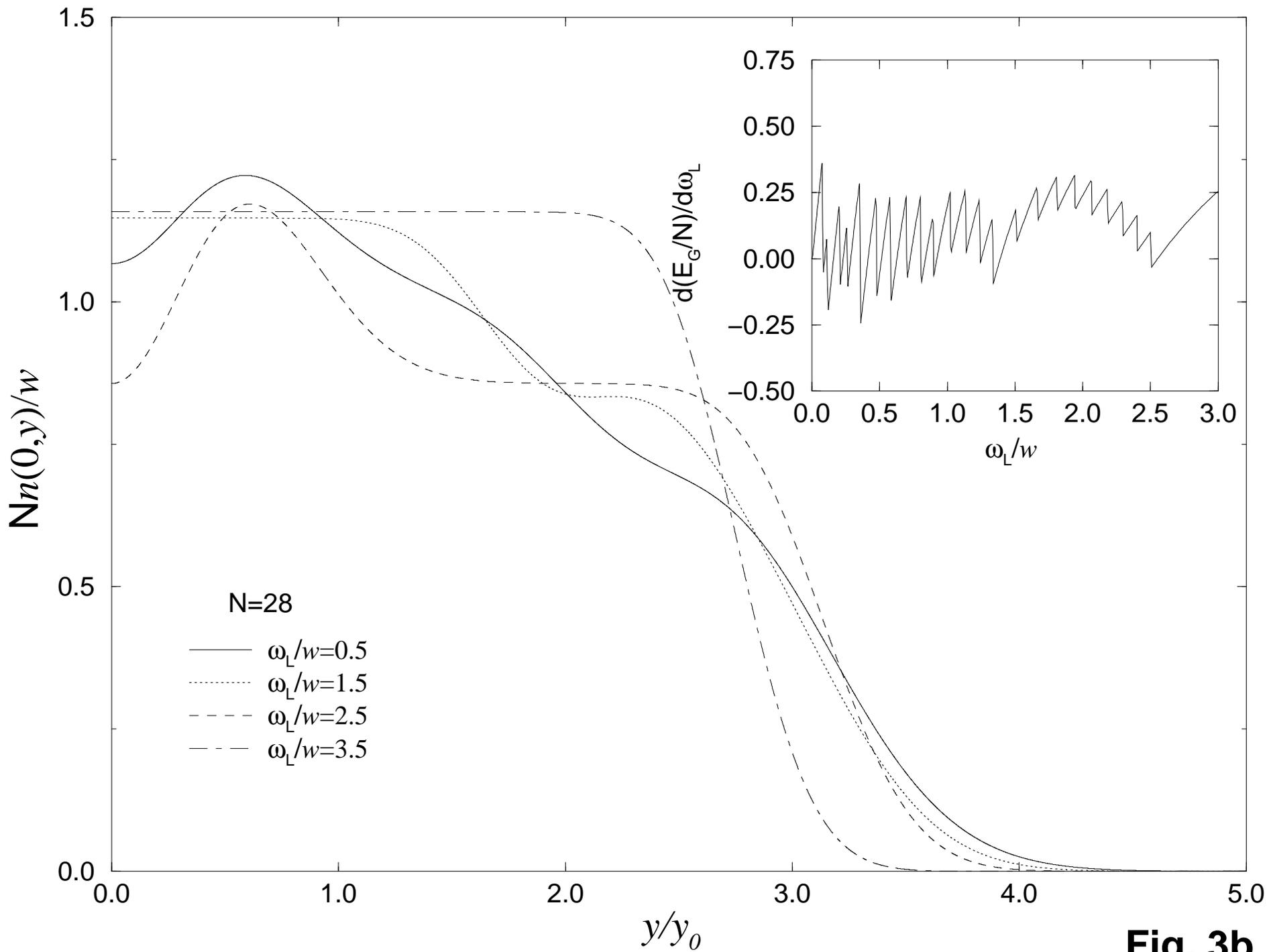
\end{document}